\documentclass{ws-rv975x65}
\usepackage{subfigure}   
\usepackage{ws-rv-thm}   
\usepackage{ws-rv-van}   
\makeindex
\def\mathnew{\mathsurround=0pt}
\def\simov#1#2{\lower .5pt\vbox{\baselineskip0pt \lineskip-.5pt
       \ialign{$\mathnew#1\hfil##\hfil$\crcr#2\crcr\sim\crcr}}}
\def\simg{\mathrel{\mathpalette\simov >}}
\def\siml{\mathrel{\mathpalette\simov <}}

\def\msun{M_\odot}

\def\beq{\begin{equation}}
\def\enq{\end{equation}}
\def\bea{\begin{eqnarray}}
\def\ena{\end{eqnarray}}
\def\bec{\begin{center}}
\def\enc{\end{center}}

\def\blist{\begin{list}{$\bullet$}{\itemsep 0.0in \parsep 0.0in}}
\def\elist{\end{list}}
\def\bitem{\begin{list}{\arabic{enumi}.}{\usecounter{enumi} \itemsep 0.0in \parsep 0.0in}}
\def\eitem{\end{list}}
\def\cm{\hbox{~cm}}
\def\s{\hbox{~s}}
\def\erg{\hbox{~erg}}

\def\MeV{\hbox{~MeV}}

\def\part{\partial}

\def\m-pl{m_{Pl}}

\def\h75{h_{75}}
\def\Omh75{\Omega h^2_{75}}
\def\Omh70{\Omega h^2_{70}}

\def\fun#1#2{\lower3.6pt\vbox{\baselineskip0pt\lineskip.9pt
  \ialign{$\mathsurround=0pt#1\hfil##\hfil$\crcr#2\crcr\sim\crcr}}}

\def\jcap{Jour. Cosmology and Astro-Particle Phys.\,}
\def\mnras{M.N.R.A.S.\,}

\def\apj{Astrophys.J.\,}
\def\apjl{Astrophys.J.Lett.\,}

\def\nat{Nature\,}
\def\na{New Ast.\,}

\def\prd{Phys.Rev.D\,}
\def\araa{Annu.Rev.Astron.Astrophys.\,}

\begin{document}

\chapter[Gamma Ray Bursts as Neutrino Sources]{Gamma Ray Bursts as Neutrino Sources}\label{ch-grb}

\author[P. M\'esz\'aros]{P. M\'esz\'aros\footnote{To appear in ``Neutrino Astronomy- Current status, future prospects", Eds. T. Gaisser \& A. Karle (World Scientific)}}

\address{Center for Particle and Gravitational Astrophysics,\\
Dept. of Astronomy \& Astrophysics and Dept. of Physics,\\
Pennsylvania State University, University Park, PA 16802, USA,\\
nnp@psu.edu\footnote{Affiliation footnote.}}

\begin{abstract}
Gamma-ray burst sources appear to fulfill all the conditions for being efficient cosmic ray
accelerators, and being extremely compact, are also expected to produce multi-GeV to PeV neutrinos.
I review the basic model predictions for the expected neutrino fluxes in classical GRBs as
well as in low luminosity and choked bursts, discussing the recent IceCube observational
constraints and implications from the observed diffuse neutrino flux.
\end{abstract}
\body

\section{Introduction}
\label{sec:intro}

Gamma-Ray Bursts (GRBs) have been postulated to be sources of very high energy (TeV to PeV)  
neutrinos since at least 1997 \cite{Waxman+97grbnu}. This is based on the 
realization that these objects may be good sites for accelerating ultra-high energy cosmic rays
(UHECRs, \cite{Waxman95cr, Vietri95cr,Milgrom+95cr}), while having an extremely high photon
luminosity which provides ideal target photons for photohadronic interactions. The observed
electromagnetic radiation is typically interpreted in terms of shock-accelerated relativistic
electrons undergoing synchrotron and inverse Compton losses. The standard fireball shock model 
of GRBs (e.g. \cite{Meszaros06grbrev}) leads to estimates for the shock region size $R$, 
comoving magnetic field strength $B'$ and comoving photon density $n'_\gamma$ which  provide
the basis for arguing that GRBs should also be sources of both UHECRs and very high energy neutrinos.
The prediction that GRBs could be strong neutrino sources has served as one of the science goals
motivating the building of the IceCube, ANTARES and the planned KM3NeT Cherenkov neutrino detectors.

\section{GRB model and variants}
\label{sec:grbmod}

GRBs are thought to be caused by a cataclysmic event at the end of the life cycle of some massive stars,
such as the collapse of the fast-rotating central core of stars more massive than $\sim 28\msun$
(giving rise to so-called ``long GRBs of gamma-ray durations $\simg$ few seconds); or the merger of 
two neutron stars or a neutron star and a stellar mass black hole, which themselves resulted from 
the previous core collapse of somewhat less massive stars (giving rise to so-called ``short GRBs, 
gamma-ray durations $\siml$ few seconds)\cite{Gehrels+09araa,Woosley+06araa}. This is the central
engine which can provide the huge energy needed to power the GRB emission which, while lasting only 
seconds, equals roughly the total luminous output emitted by the Sun over $10^{10}$ years, or that 
emitted by the entire galaxy over a hundred years, and is detectable out to the farthest reaches 
of the Universe.

The collapse or merger results in the liberation of a gravitational energy of order $E_{grav} \sim 
GM^2/r \sim 10^{54}\erg$ on a very short timescale, in a region whose dimensions $r_0$ are of order 
of tens of kilometers, leading to a fireball of photons, $e^\pm$ pairs, magnetic fields and 
baryons \footnote{Most of the liberated $E_{grav}\sim 10^{54}\erg$, however, escapes as a $\sim 10\s$ 
burst of $\sim 10-30\MeV$ thermal neutrinos, as in supernovae, and as gravitational waves.}.
This fireball, which is initially extremely optically thick, expands most easily 
along the rotation axis, driven by the radiation pressure. The expansion is highly relativistic, 
characterized by bulk Lorentz factors of the fireball plasma of order $\Gamma\sim 10^2-10^3$, as 
inferred from the observation of multi-GeV photons \cite{Meszaros06grbrev}. This requires the fireball
to have a small baryon load $M_{0}c^2$ compared to the fireball energy $E_{0}$, i.e. a high dimensionless 
entropy $\eta=E_{0}/M_{0} c^2 \sim 10^2-10^3$. The actual outflow is inferred, from observations of the 
light-curves \cite{Gehrels+09araa} to be collimated into jets of opening angle $\theta_j \sim 0.1$, 
which reduces the fireball energy requirements to ${\cal O}(10^{51}\erg)$

In the standard fireball model, if the inertia is dominated by the baryon load, the Lorentz
factor grows as $\Gamma\propto r$ by converting internal into bulk kinetic energy, up to 
$\Gamma_f\simeq \eta\simeq$ constant, at $r_s \simg r_0\eta$ after which the ejecta coasts
\cite{Paczynski86,Goodman86grb,Shemi+90}.
Beyond a photospheric radius $r_{ph}$ where the ejecta becomes optically thin  to 
Thompson scattering the fireball quasi-thermal gamma-rays can escape freely. For $r>r_s$ however,
most of the fireball energy is kinetic, rather than in the form of photons, and the spectrum
is quasi-thermal, contrary to most of the observed burst spectra.
\begin{figure}[h]
\vspace*{-0.0in}
\begin{minipage}{0.5\textwidth}
\centerline{\includegraphics[width=3.1in,angle=0]{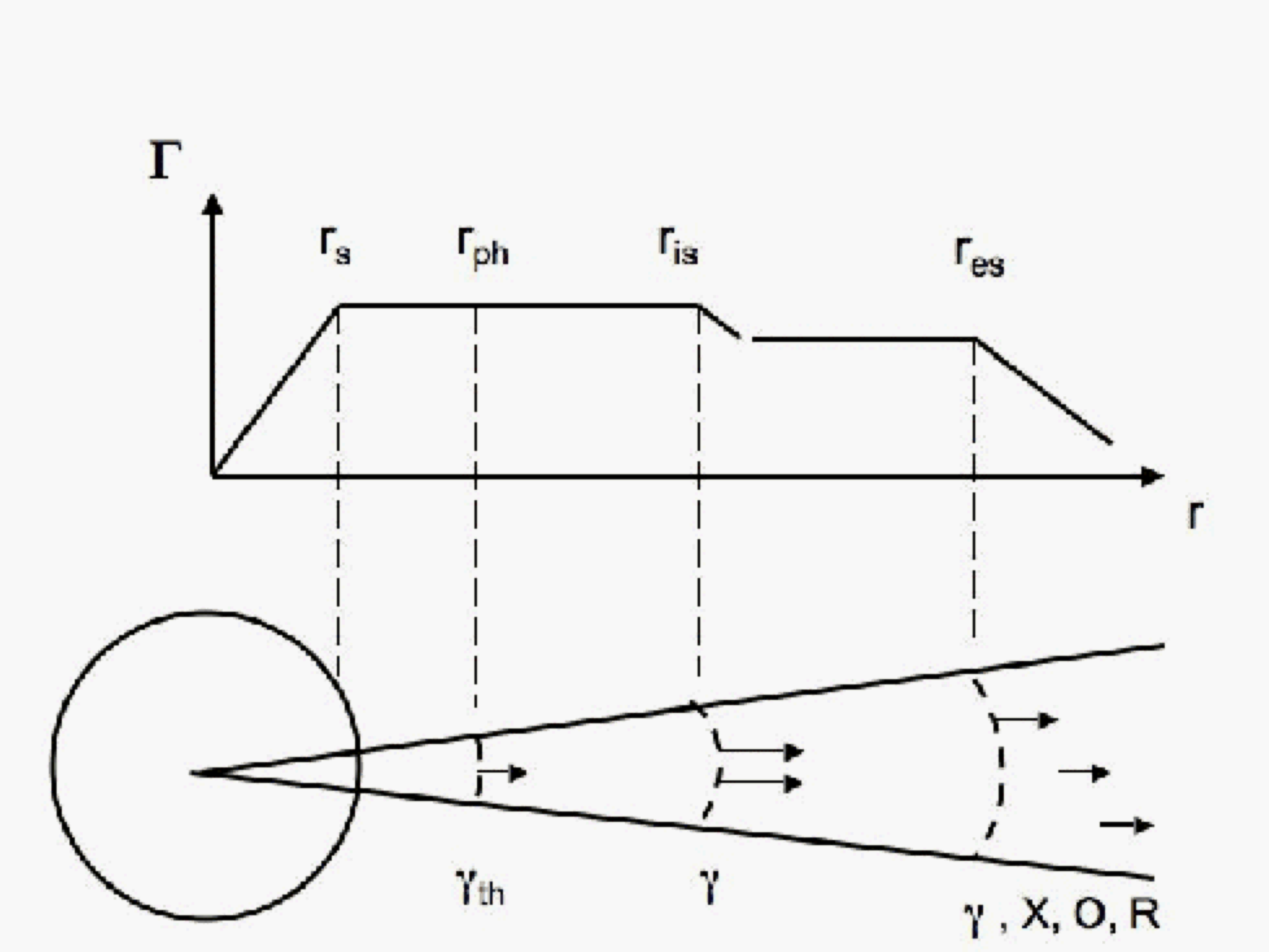}}
\end{minipage}
\hfill
\hspace{5mm}
\begin{minipage}[t]{0.40\textwidth}
\vspace*{-0.4in}
\caption{Top: schematic evolution of the bulk Lorentz factor for a GRB baryonic outflow.
Bottom: the tree main emission zones from which gamma-rays are detectable.}
\end{minipage}
\label{fig:lorentz}
\end{figure}

The most widely accepted paradigm for producing the $\gamma$-ray spectrum seen in the majority of 
GRBs is referred to as the standard fireball shock model, which naturally produces a non-thermal
spectrum, and also increases the efficiency by tapping the large reservoir of expansion kinetic energy 
\cite{Rees+92fbal,Meszaros+93impact,Rees+94unsteady}.
Collisionless shocks are expected outside the photosphere, where the ejecta is optically thin, leading 
to Fermi acceleration of particles to a relativistic power law distribution, leading to 
broken power-law synchrotron and inverse Compton spectra.
Two types of shocks are expected: internal shocks at $r_{is}\simg r_{ph}$ caused by variations in  the 
ejecta Lorentz factor \cite{Rees+94unsteady}, which can give rise to the fast-varying prompt non-thermal 
$\gamma$-ray emission;  and external shocks at $r_{es} \simg r_{is} \simg r_{ph}$ \cite{Rees+92fbal}, 
giving rise to the longer-lasting X-ray, optical and radio afterglows \cite{Meszaros+97ag}.
The typical radii of the photosphere (for a baryon dominated ejecta) and the shocks are 
\bea
r_{ph} &\simeq (L_0\sigma_T /4\pi m_pc^3\eta^3)
   \sim 4\times 10^{12} L_{\gamma,52} \eta_{2.5}^{-3}\cm\cr
r_{is}& \simeq  \Gamma^2 c t_v
  \sim 3\times 10^{13} \eta_{2.5}^2 t_{v,-2}\cm~~~~~~~~~~~~~~~~~~~\cr
r_{es}& \simeq (3E_0/4\pi n_{ext}m_pc^2\eta^2)^{1/3} ~~~~~~~~~~~~~~~~~~~~~~~~~~~~\cr
  & \sim 2 \times 10^{17}(E_{53}/n_0)^{1/2}\eta_2^{2/3}\cm ,~~~~~~~~~~~~~~.
\label{eq:3radii}
\ena
Here we used $E_0,L_0,\eta\sim\Gamma,n_{ext}, t_v$ as the burst total energy, luminosity, initial
dimensionless entropy, coasting bulk Lorentz factor, external density and intrinsic time variability
\cite{Zhang+04grbrev,Meszaros06grbrev}.
\begin{figure}[h]
\vspace*{-0.0in}
\begin{minipage}{0.5\textwidth}
\centerline{\includegraphics[width=3.1in,angle=0]{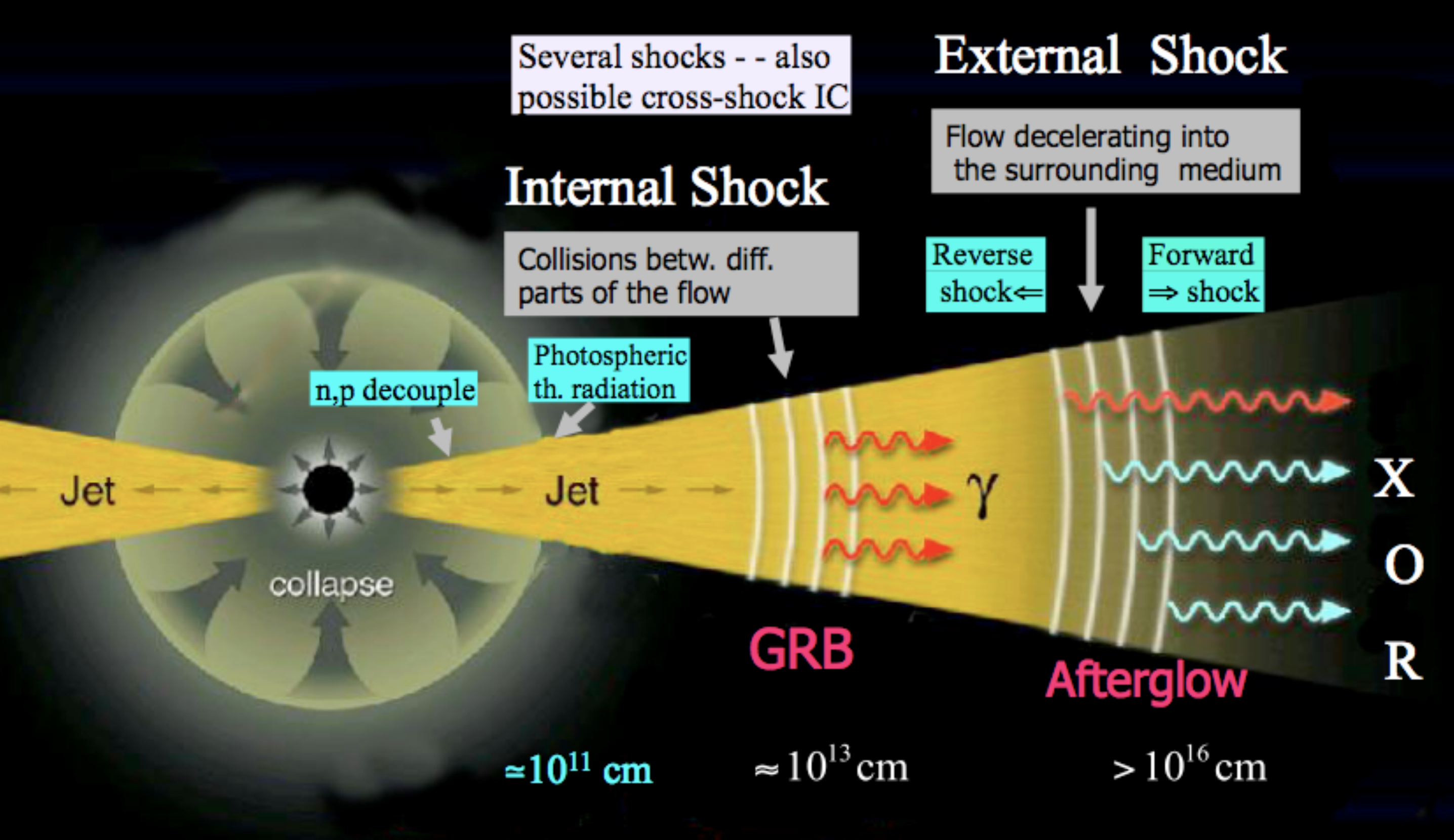}}
\end{minipage}
\hfill
\hspace{5mm}
\begin{minipage}[t]{0.3\textwidth}
\vspace*{-0.4in}
\caption{Schematic GRB jet emission zones (either or both $\nu$ and $\gamma$), starting with 
sub-photospheric (innermost) to photosphere to internal shock (IS) top external shock (ES).}
\end{minipage}
\label{fig:jet-schem}
\end{figure}

The prompt $\gamma$-ray spectra are usually phenomenologically fitted with a ``Band" broken 
power-law spectrum. In some bursts the fitted low energy spectral slopes appeared initially
to be harder then $\alpha >-2/3$, which would violate the low energy asymptote of a synchrotron 
spectrum, e.g. \cite{Meszaros06grbrev}. To address this issue, a combination of a passive 
photospheric quasi-blackbody spectrum at low energies and a power-law shock synchrotron
spectrum at high energies were considered \cite{Meszaros+00phot}, and in fact, evidence for low 
energy quasi-thermal emission compatible with a photosphere has been detected in a number of bursts 
\cite{Ryde+06phot,Ryde+11phot}.
More recently, it has been confirmed \cite{Guiriec+15-LEpeak,Guiriec+15thermnth}
that considering the joint effects of a photosphere-like quasi-thermal spectrum together with a
Band broken power-law non-thermal spectrum results in fitted slopes which are compatible with 
a synchrotron interpretation. The latter could be a shock outside the photosphere.

An alternative view of the origin of the prompt $\gamma$-ray emission, which similarly addresses
the low energy slope issue and in addition also the efficiency issue, is that the entire prompt
emission arises in a dissipative photosphere \cite{Rees+05photdis} (as opposed to a passive, 
adiabatic photosphere). The low energy slope is hard because it is self-absorbed, while
the high energy slope is a power law due to comptonization, e.g. \cite{Peer+06phot}. 
The dissipation could be due to internal shocks at or below the photosphere, or else it  could be
due to magnetic field reconnection \cite{Meszaros+11gevmag} or it could be due to collisional effects  
following the decoupling of protons and neutrons below the photosphere \cite{Beloborodov10pn}.
In this case the high energy photon power law slope extension is due to upscattering of the thermal
photons by relativistic positrons from pion decay following $pn$ collisions.

\section{VHE neutrinos from GRBs}
\label{sec:nugrb}

The co-acceleration of ions is natural in models where electrons are accelerated (as inferred
from the gamma-ray observations), if besides electrons the ejecta contains also baryons. The
latter is expected in a stellar core collapse or merger event, where the mass density is close
to nuclear. The detailed model fits indicate that the baryon load is small but non-negligible,
as inferred from termination bulk Lorentz factors $\Gamma_f \sim E_0/M_0c^2 \sim 10^2-10^3$.
In such scenarios, VHE neutrino production in the shock or acceleration zones is expected from $p\gamma$
interactions. Other scenarios where the stress-energy is dominated by $e^\pm$ or magnetic fields 
(which could have much fewer or no baryons, as in pulsar models) are possible,  and would imply
negligible neutrino production. However, such models would naturally be associated with much larger 
bulk Lorentz factors than those inferred from observations. In such models the flow further out 
may decelerate due to pair drag \cite{Meszaros+97poynt} or it may pick up more baryons, but
such scenarios involve more free parameters and are not widely considered in model fits.
Models where the stress-energy is largely magnetic which do have small but appreciable baryon 
loads have been considered, e.g. \cite{Tchekhovskoy+10grb,Metzger+11grbmag}, which lead to 
observationally acceptable final Lorentz factors, and having baryons, also fulfill a necessary
condition for being potential neutrino sources.

The most straightforward prediction for VHE neutrino production is that associated with internal 
shocks \cite{Waxman+97grbnu}. This was initially based on a simplified internal shock, with given
fixed shock radius parameterized by the total gamma-ray energy, Lorentz factor and outflow time 
variability, and approximating the photon spectrum as an average-slope Band broken power law,
using the $\Delta$ resonance approximation for the photohadronic interaction. This simplified
model was adopted \cite{Guetta+04grbnu}  to make the first predictions of an expected  diffuse
VHE neutrino background, assuming a relativistic proton to electron luminosity ratio $f_p=f_e^{-1}
=L_p/L_e$ and taking $L_\gamma\simeq L_e$, for a given a set of electromagnetically observed bursts.
Using 215 bursts with known $\gamma$-ray fluences, the first IceCube observations with 40 strings 
and later 56 strings were compared \cite{Ahlers+11-grbprob,Abbasi+11-ic40nugrb,Abbasi+12grbnu-nat}
to the diffuse flux predicted by this a simplified internal shock (IS) model. They concluded that
for a nominal $f_p=f_e^{-1}=L_p/L_e=10$ ratio the model over-predicted the data by a factor 5, and a
model-independent analysis comparing the observed diffuse neutrino flux to that expected if GRBs 
were the sources of the observed UHECR flux was also similarly off. This was a very important 
first result using a major Cherenkov neutrino facility to constrain astrophysical source models.

Subsequent analyses pointed out {\cite{Li12grbnu,Hummer+12nu-ic3} that using the same fixed shock 
radius IS model but correcting for various approximations and including besides the $\Delta$-resonance 
also  multi-pion and Kaon channels as well as interactions with the entire target photon spectrum,
lower predicted fluxes are obtained which not disagree with the 40+56 string data, and which indicate 
that 5 years of observations might be needed with the full 86 string array to rule out the simple IS model. 

An issue with the fixed-radius IS models is that, even if one uses a distribution of radii
$r_{is}$ in eq.(\ref{eq:3radii}) based on observational distributions of variability times 
$t_v$ and bulk Lorentz factors $\Gamma$, this still denotes the shock initiation radius, and the 
baryon, photon and magnetic field densities decrease as the shocked mass shells expand beyond
this radius, necessitating a time dependent calculation in addition to a statistical averaging.
Such time-dependent IS proton acceleration and full physics $p\gamma$ neutrino diffuse flux
calculations \cite{Asano+14grbcr,Bustamante+15grbcrnu} result, as expected, in a diminished 
predicted neutrino flux, as the $p\gamma$ opacity drops with distance away from the initial radii, 
giving a result which is well within the bounds of the \cite{Abbasi+12grbnu-nat} 40+56 string 
upper limits. It remains to redo such comparisons against the full array data.
\begin{figure}[h]
\vspace*{-0.0in}
\begin{minipage}{0.5\textwidth}
\centerline{\includegraphics[width=3.1in,height=2.5in,angle=0]{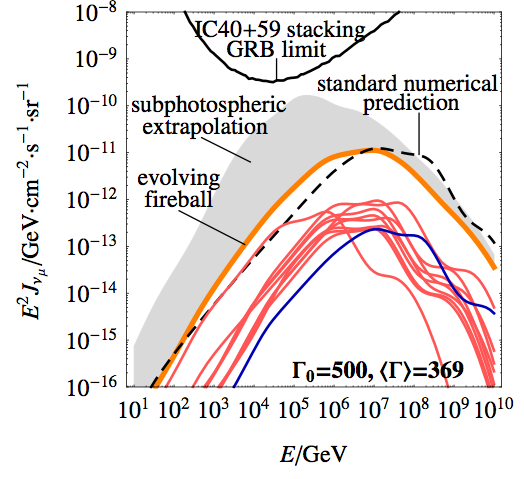}}
\end{minipage}
\hfill
\hspace{5mm}
\begin{minipage}[t]{0.4\textwidth}
\vspace*{-0.4in}
\caption{Results from a time-dependent internal shock 
neutrino production calculation, including a sub-photospheric 
contribution, compared to IceCube 40+59. From \cite{Bustamante+15grbcrnu}.}
\end{minipage}
\label{fig-bustam15}
\end{figure}

Of course, an open question with GRBs is whether the basic internal shock model is correct for the 
prompt gamma-ray emission (and its related proton acceleration and $p\gamma$ production).  The 
$\gamma$-ray spectral issues may no longer be a concern for IS models \cite{Guiriec+15-LEpeak,
Guiriec+15thermnth} but the mechanical radiative efficiency still remains a question. On the other 
hand, dissipative photospheric models of the prompt $\gamma$-ray emission appear to address both 
the efficiency and spectra adequately, e.g. \cite{Beloborodov10pn,Meszaros+11gevmag,Zhang+11icmart}. 
Early diffuse neutrino flux predictions from baryonic  GRB photospheres were presented in
\cite{Murase08grbphotnu,Wang+09grbphotnu}. The results are different for magnetically dominated 
outflows, where the initial acceleration can be parameterized through $\Gamma(r)\propto r^\alpha$ 
where $1/3 < \alpha <1$, as opposed to $\Gamma(r)\propto r$ in the baryon-dominated case.
Diffuse neutrino flux predictions from both baryon-dominated and magnetically dominated photospheres 
were calculated by \cite{Gao+12photnu,Gao+12magnu,Bartos+13pn} indicating no violation of the 40+56 
string IceCube constraints, but approaching it. The main reason is that the photosphere radii are 
smaller than those of internal shocks, see eq.(\ref{eq:3radii}), hence photon and particle densities and 
$p\gamma$ optical depths are larger. Model independent calculations confirm this \cite{Zhang+13grbnu}. 

A more recent analysis of the IceCube four years data, including two years of the full array 
\cite{IC3+15grbnu4yr}, used the full physics of the $p\gamma$ interactions and the entire 
target photon spectrum to compare against the predictions of the standard IS Model, baryonic 
photosphere model and ICMART model, all three for fixed radius (steady state) emission zones.
\begin{figure}
\centerline{\includegraphics[width=10cm]{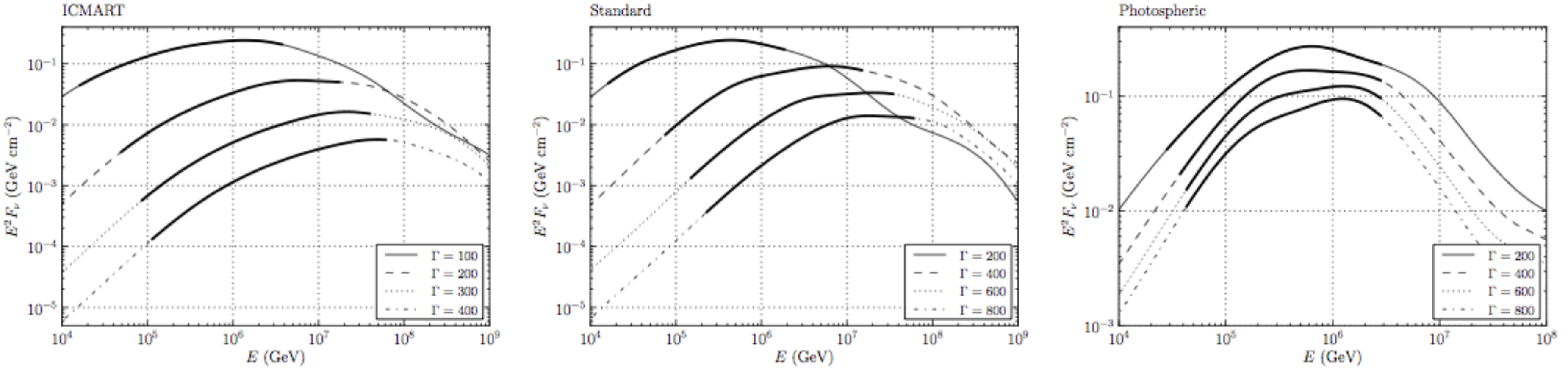}}
\centerline{\includegraphics[width=10cm]{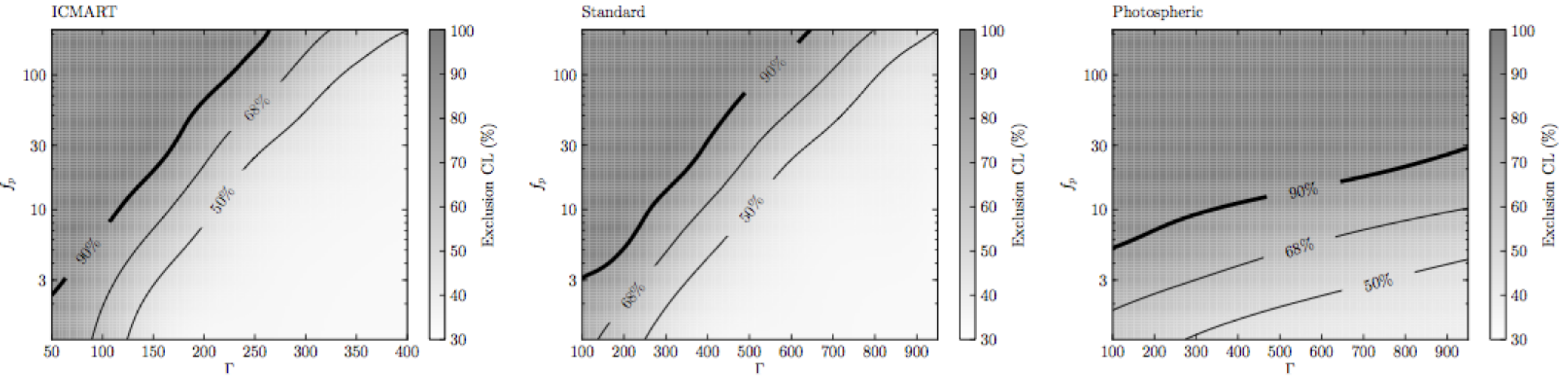}}
\caption{Top: Total normalized neutrino fluxes for ICMART, IS and (baryonic) photosphere models 
(left to right) for various $\Gamma$, scaling with $f_p$ (which here is 10).
Bottom: Allowed region for $f_p$ and $\Gamma$ for the different models. From \cite{IC3+15grbnu4yr}.}
\label{ic3+15grb3lim}
\end{figure}
They concluded that at 99\% confidence level less than 1\% of the observed diffuse neutrino 
background can be contributed by the observed sample of 592 electromagnetically detected GRBs.
This is a much larger burst sample with a more complete array,  and while it would be important 
to redo this analysis with time-dependent expanding radius emission zones, this is likely to be
a much stronger constraint. If the basic acceleration paradigm in these emission zones is correct,
and the result continues to stand, it may be indicating that the ratio $L_p/L_e=f_p \siml 1$.
Other photospheric models with substantially different neutrino production physics 
\cite{Murase+13subphotnu,Kashiyama+13pnconv} 
have been investigated, but so far have been only qualitatively compared against the data.

\section{Other types of GRBs as possible neutrino sources}
\label{sec:alt}

The classical GRBs discussed above are what may be called ``overt" (electromagnetically detected) GRBs,
the majority (70-90\%) of which are long GRBs ascribed to core-collapse events located at redshifts 
$z\simg 1$. The rest, about 30-10\% of the classical GRBs are short GRBs, which appear to be compact 
mergers and which have a luminosity lower by about one order of magnitude, being detected typically 
at lower redshifts $z\siml 1-1.5$. Their neutrino luminosity probably scales with their gamma-ray 
luminosity, and as such they are not expected to contribute much to the long GRB predicted diffuse 
neutrino fluxes.

There are, however, at least three other classes of GRBs in addition to the classical ones, which 
could contribute to the diffuse neutrino flux. These are the low-luminosity GRBs (LLGRBs), the
choked GRBs, and the shock break-out GRBs (which may be an intermediate or transition class between 
the LLGRBs and the choked GRBs). 

The low luminosity GRBs, not surprisingly, have been discovered only at low redshifts, some as
low as $z =0.0085$ (GRB980425/SN1998bw); at the same time, because of the low redshift, in most
of these an associated supernova of type Ic has been spectroscopically detected. While the total
number of detected LLGRBs is only a handful, the inferred local rate appears to be about an order 
of magnitude higher (per unit volume and time) than that of classical GRBs, e.g. 
\cite{Howell+13-llgrb,Virgili+09llgrb,Soderberg+06-060218}. A simple scaling of the classical 
long GRB IS shock paradigm has led to the expectation that LLGRBs could contribute a significant, 
or perhaps even dominant fraction of the total GRB UHECR and VHE neutrino diffuse fluxes 
\cite{Murase+06llgrbnu,Murase+08llgrbcr,Gupta+07nullgrb,Liu+11-llgrbcr}. 

Choked GRBs are core collapse objects similar in their dynamics to the observed classical
long GRBs, where a relativistic jet has been launched from a central engine, the difference being
that the jet did not make it out from the star, having stalled either because the accretion
onto the central object did not last long enough, or because the stellar envelope is larger
than in ``successful" GRBs, where the jet has emerged \cite{Meszaros+01choked}. Internal
shocks can be expected in such jets while they are still below the stellar surface, which can 
also accelerate protons and undergo $p\gamma$ interactions leading to neutrinos that emerge 
\cite{Meszaros+01choked}. In successful or overt GRBs (where the jet emerges and makes $\gamma$-rays) 
this neutrino burst from the sub-stellar phase of the jet acts as a precursor, while in truly choked jets, 
where the jet stalls and which are $\gamma$-dark, the neutrino burst reveals a failed GRB, which is
a forerunner of a jet-boosted supernova. The spectrum should have a low energy (multi-GeV) $pp$ 
component as well as a higher energy $p\gamma$ TeV component, and these could be used to diagnose 
the stellar envelope  extent and structure \cite{Razzaque+04nuhn,Razzaque+03nutomo,Ando+05sngrbnu}.
More detailed calculations \cite{Horiuchi+08choked,Horiuchi+09choked,Murase+13choked} have
explored possible signatures and their potential detectability by IceCube and Deep Core,
preliminary results and limits having been presented \cite{Taboada11choked}. 

The shock which propagates through the envelope of core collapse supernovae (ccSNe) eventually breaks 
out of the envelope, and this may happen whether a jet was launched from the core or not, and whether
such a jet eventually emerges or not from the envelope and/or the optically thick precursor 
stellar wind. When it does emerge, an X-ray flash is observed; in some cases this was observed in 
what appeared to be a LLGRB, e.g. \cite{Campana+06-060218}; in another case it was seen in a core-collapse
supernova unassociated with a GRB \cite{Soderberg+08sn}. It is tempting to identify the former
with  ccSNe where a jet was launched and emerged, and the latter with ccSNE where the jet did
not emerge (or perhaps was not even launched, due to lack of enough angular momentum to feed
a central accreting black hole or magnetar). The shock break-out occurs when the photons which 
previously were diffusively trapped become able to escape  freely, which is thought to occur above 
the ejecting envelope, in the optically thick wind which precedes the SN explosion, e.g. 
\cite{Waxman+07-060218,Chevalier+08break}.  If a jet was launched, an anisotropy of the envelope
and the wind is expected, as indicated by the interpretation of GRB 060218 \cite{Waxman+07-060218}.
This would be expected whether the jet emerged or not\footnote{In fact, many SN remnants show at 
late stages optical polarization attributed to scattering by an anisotropic ejecta}. 
The ejecta of several of the SNe associated to LLGRB appear to have a semi-relativistic component
\cite{Soderberg+06-060218,Campana+06-060218}, which is interesting for the production of UHECR, 
e.g. cite{Wang+07crhn,Wang+07crhn}. The details of the shock break-out process, and whether the 
envelope becomes semi-relativistic (as appears to be the case in most of the SNe accompanying LLGRB) 
is of continued interest
especially for its impact on their possible UHECR and VHE neutrino production, e.g.
\cite{Katz+11snrnu,Murase+13subphotnu,Kashiyama+13breakout}.

\section{The TeV-PeV diffuse neutrino background}
\label{sec:nubkg}

The observed diffuse flux of sub-PeV to PeV neutrinos detected by IceCube
\cite{IC3+13pevnu1,IC3+13pevnu2} and its extension down to the TeV range \cite{IC3+15tevnu} is 
believed to be of astrophysical origin.
So far, however, no significant spatial or temporal correlations have been found with classical 
(high luminosity) GRBs detected electromagnetically by Swift or Fermi \cite{IC3+15grbnu4yr}, 
nor for that matter with any other type of known sources. Various possible types of candidate 
sources have been considered (a partial list is in \cite{Senno+15clugalnu}). In particular
starburst galaxies are a possibility (see \cite{Loeb+06nustarburst} and other references cited in
\cite{Senno+15clugalnu}), within which the actual production sites are likely to be hypernovae and
supernovae, e.g. \cite{Senno+15clugalnu}. Another possibility are low-luminosity GRBs (LLGRBs, see
\S \ref{sec:alt}) \cite{Murase+06llgrbnu,Murase+08llgrbcr,Gupta+07nullgrb,Liu+11-llgrbcr}. 
Their rate being higher than that of classical GRB, they could provide a significant neutrino background, 
and in $\gamma$-rays they are detectable only at low redshifts (a handful so far) but not at 
$z\simg 0.5-1$. However, one would expect sooner or later a $\nu-\gamma$ coincidence at low redshifts. 
Another interesting possibility are the choked GRBs \cite{Meszaros+01choked,Horiuchi+08choked,
Horiuchi+09choked,Murase+13choked,Fraija15pevchoked}. The shocks accelerating protons occur in 
the jet inside the star, hence they would be $\gamma$-dark, although the ejected envelope could 
lead to longer optical/IR longer transients or supernovae. Note that buried jets of arbitrary luminosity,
e.g. \cite{Fraija15pevchoked}, are subject to the caveat \cite{Murase+13choked} that high (classical) 
luminosities lead to radiation dominated buried shocks, which prevents Fermi acceleration; 
collisionless shocks able to Fermi-accelerate protons are expected only for low luminosity choked jets. 
Both LLGRBs and choked jets, being associated to ccSNe, are also likely to be predominantly located 
in starburst or starforming galaxies. 
\\

\noindent
{Partial support from NASA NNX13AH50G as well as useful discussions with K. Murase and N. Senno are
gratefully acknowledged.}


\end{document}